\documentclass[11pt,a4paper]{article}
\pdfoutput=1

\usepackage[intlimits]{amsmath}
\usepackage{amsfonts}
\usepackage{amssymb}
\usepackage{mathrsfs}
\usepackage[latin1]{inputenc}
\usepackage[T1]{fontenc}
\usepackage{graphicx}
\usepackage{epstopdf}
\usepackage{authblk}
\usepackage[english]{babel}

\def\a{\alpha}
\def\b{\beta}

\def\G{\Gamma}
\def\d{\delta}

\def\z{\zeta}

\def\La{\Lambda}

\def\r{\rho}

\def\s{\sigma}

\def\t{\tau}
\def\f{\phi}

\def\vf{\varphi}

\def\ps{\psi}
\def\o{\omega}

\newcommand{\ti}[1]{\tilde{#1}}

\newcommand{\ha}{\hat{a}}

\providecommand{\babs}[1]{\big\lvert#1\big\rvert}
\providecommand{\Babs}[1]{\Big\lvert#1\Big\rvert}
\providecommand{\bbabs}[1]{\bigg\lvert#1\bigg\rvert}

\providecommand{\norm}[1]{\lVert#1\rVert}

\newcommand{\de}{\partial}
\newcommand{\p}{\prime}
\newcommand{\nn}{\nonumber}
\newcommand{\vecx}{\mathbf{x}}
\newcommand{\vecy}{\mathbf{y}}
\newcommand{\vecz}{\mathbf{z}}
\newcommand{\veck}{\mathbf{k}}
\newcommand{\dg}{\dagger}

\newcommand{\mcalB}{\mathcal{B}}
\newcommand{\mcalR}{\mathcal{R}}
\newcommand{\mcalH}{\mathcal{H}}
\newcommand{\mscrP}{\mathscr{P}}

\newcommand{\mscrS}{\mathscr{S}}
\newcommand{\mscrH}{\mathscr{H}}
\newcommand{\mbbN}{\mathbb{N}}
\newcommand{\mbbZ}{\mathbb{Z}}
\newcommand{\mbbR}{\mathbb{R}}

\numberwithin{equation}{section}

\begin{document}

\title{The divergence of Van Hove's model and its consequences}

\author{\Large{Fulvio Sbis\`a\footnote{fulviosbisa@gmail.com ; https://orcid.org/0000-0002-6341-1785 .}}}

\affil{\normalsize{Departamento de F\'isica Te\'orica, Universidade do Estado do Rio de Janeiro,\\
CEP 20550-013, Rio de Janeiro -- RJ -- Brazil}}

\date{}

\vspace{.5cm}

\maketitle

\thispagestyle{empty}

\begin{abstract}
We study a regularized version of Van Hove's 1952 model, in which a quantum f\mbox{}ield interacts linearly with sources of f\mbox{}inite width lying at f\mbox{}ixed positions. We show that the central result of Van Hove's 1952 paper on the foundations of Quantum Field Theory, the orthogonality between the spaces of state vectors which correspond to dif\mbox{}ferent values of the parameters of the theory, disappears when a well-def\mbox{}ined model is considered. We comment on the implications of our results for the contemporary relevance of Van Hove's article.
\end{abstract}


\section{Introduction}

The existence of Unitarily Inequivalent Representations (UIR) of the Canonical Commutation Relations (CCR) is an unavoidable feature of quantum systems with an inf\mbox{}inite number of degrees of freedom, since in that case the Stone-von Neumann theorem \cite{Stone 1930,von Neumann 1931} does not hold. Although their existence were known at least before World War II \cite{von Neumann Comp Math 1939} (to von Neumann, at any rate), the community of physicists started to become more widely aware of their relevance mainly due to the works of Van Hove \cite{Van Hove 1952} and Friedrichs \cite{Friedrichs 1952a,Friedrichs 1952b}. In particular, Van Hove proposed a simple model where the existence of UIR was revealed by the circumstance that the spaces of state vectors relative to dif\mbox{}ferent values of the model's parameters are orthogonal. This showed how (borrowing Wightman and Schweber's words \cite{Wightman Schweber 1955}) ``other representations of the commutation rules are not pathological phenomena whose construction requires mathematical trickery, since they occur in the most elementary examples of f\mbox{}ield theory''. See \cite{Lupher 2005,Earman Fraser 2006} for a historical recollection and for a discussion of the connection of Van Hove's model with Haag's theorem \cite{Haag 1955,Hall Wightman 1957}.

The model considered by Van Hove is apparently innocent, consisting of several point-like charges lying at f\mbox{}ixed positions which interact with a neutral and relativistic massive scalar f\mbox{}ield. The whole analysis rests on the remarkable fact that the quantum equations of motions can be solved \emph{exactly}, without resorting to the perturbative approach. While acknowledging that the model is unrealistic, Van Hove conjectured that the spaces of state vectors relative to dif\mbox{}ferent values of the coupling constant would remain orthogonal, or at least dif\mbox{}ferent, also in a more realistic QFT. From a mathematical point of view, however, the model is singular, since the exact solutions are divergent. Indeed the analysis becomes possible only after neglecting an inf\mbox{}inite contribution to the ground state energy of the system (\emph{besides} the already discarded ground state energy of the quantum harmonic oscillators), operation which is in no way justif\mbox{}ied in the article.

In light of its historical relevance, our aim in this paper is to perform a critical scrutiny of the article's analysis. Two points in particular are examined: the f\mbox{}irst is whether the phenomenon of orthogonality of the spaces of state vectors would still be present if we considered a less unrealistic modelization of the sources, such as one where the sources have f\mbox{}inite but non-zero radius or a charge prof\mbox{}ile rapidly decaying at inf\mbox{}inity. The second, related, point is whether it is possible to modify the model in such a way to regularize the above mentioned divergence while preserving the non-trivial phenomenology. A strong correlation between the divergence and the orthogonality phenomenon, in fact, would raise the suspicion that the latter is just an artifact of an ill-based model. Conversely, the absence of such a correlation would provide an a posteriori justif\mbox{}ication for the procedure of neglecting the inf\mbox{}inite energy contribution. An important part of this investigation is to understand the physical meaning of the divergence.

The paper is structured as follows. In section \ref{sec: More realistic model} we consider a regularized version of Van Hove's model and investigate the physical meaning of the above mentioned divergence. The orthogonality of the spaces of state vectors in the regularized model is investigated in section \ref{sec: Orthogonality spaces state vectors}, and the implications of our results are discussed in section \ref{sec: Discussion}. We gather our conclusions in section \ref{sec: Conclusions}.

\section{The model and its generalization}
\label{sec: More realistic model}

Although in the original model \cite{Van Hove 1952} the scalar f\mbox{}ield interacts with an arbitrary (f\mbox{}inite) number of sources, we focus here on the special case of two sources, which we believe to be suf\mbox{}f\mbox{}iciently representative of the general situation. We refer to the recent exposition \cite{Sbisa 2020} of Van Hove's article for a detailed reference.

To recall its main features, the model consists of two point-like sources placed respectively at the positions $\vecy_{_{\!1}}\!$ and $\vecy_{_{\!2}}\!$ (with $\vecy_{_{\!1\!}} \neq \vecy_{_{\!2\!}}$ strictly), interacting with a real, relativistic scalar f\mbox{}ield $\f(t,\vecx)$ of mass $m$ via an interaction term linear in $\f$ and proportional to the charge of the source. The system is enclosed in a cubic box of large volume with periodic boundary conditions, so the f\mbox{}ield belongs to the space of (square-integrable) functions periodic on a three-dimensional cubic lattice. We indicate with $L$ the length of the edges of the primitive cell $\mscrP$, and $V = L^{3}$ indicates its volume, so the reciprocal lattice, which we indicate with $\mcalR\,$, is again a cubic lattice whose primitive cell has edges of length $2 \pi/ L$ and volume equal to $(2 \pi)^{3}\!/ V$.

\subsection{Spatially distributed sources}

As anticipated above, the model is idealized in various ways. For one thing, the sources create the f\mbox{}ield but don't react to it, their positions being f\mbox{}ixed (``inf\mbox{}inite mass'' approximation). Besides, they have no internal structure and have zero spatial width, being point-like. The second property seems the most delicate, in light of the fact that strictly speaking it is not consistent to def\mbox{}ine a quantum f\mbox{}ield sharply at a point \cite{Streater Wightman 1964}.

We therefore consider a slightly less idealized model, in which the sources have a f\mbox{}inite spatial width and an internal structure. This can be done easily by introducing two smooth (real) functions $\r_{_{1}}\!$ and $\r_{_{2}}\!$ of compact support, normalized to one and centered around the origin. The idea is that the product $q_{i} \, \r_{i}\,$, suitably shifted in space, plays the role of charge density distribution of the $i$-th source. We assume that the diameter of the supports of $\r_{_{1\!}}$ and $\r_{_{2\!}}$ is much smaller than the length $\norm{\vecy_{_{\!1}}\! - \vecy_{_{\!2\!}}}\,$, and that their supports remain contained in the interior of $\mscrP$ when translated by the vectors $\vecy_{_{\!1}}\!$ and $\vecy_{_{\!2}}$. We then consider the ref\mbox{}ined model given by
\begin{align}
H &= H_{_{0}} + g H_{\textup{i}} \quad , \label{H tot} \\[6mm]
H_{_{0}} &= \frac{1}{2} \, \int_{\mscrP} \! \Big( \pi^{2} + \babs{\vec{\nabla} \f}^{2} + m^{2} \f^{2} \Big) \, d^{3} x \quad , \label{H 0} \\[1mm]
H_{\textup{i}} &= \sum_{i = 1}^{2} \, q_{i} \int_{\mscrP} \! \r_{i}(\vecx - \vecy_{\!i}) \,\, \f(t, \vecx) \,\, d^{3}x \quad . \label{H i}
\end{align}
where the coupling constant $g$ is adimensional and natural units $\hbar = c = 1$ are used. The charge density distributions are f\mbox{}ixed, which means that both the position of the sources and their internal structure still do not react to the f\mbox{}ield. The description is still quite idealized, but slightly less unrealistic. Naively speaking, Van Hove's model is reproduced when the functions $\r_{_{1\!}}$ and $\r_{_{2\!}}$ ``tend to a Dirac delta''.

\subsection{Quantization}

The quantization proceeds exactly as in \cite{Van Hove 1952}. We keep the derivations short and refer to \cite{Sbisa 2020} for a more detailed exposition. Upon canonically quantizing the model (\ref{H tot})--(\ref{H i}) in the Schr\"odinger picture, the f\mbox{}ield operator and its canonically conjugate momentum read
\begin{align}
\hat{\phi}(\vecx) &= \frac{1}{\sqrt{2V}} \, \sum_{\veck \in \mcalR} \frac{1}{\sqrt{\o_{k}}} \, \Big( \ha^{\phantom{\dagger}}_{\veck} \, e^{i \veck \cdot \vecx} + \ha^{\dagger}_{\veck} \, e^{- i \veck \cdot \vecx} \Big) \label{quantised phi} \\[2mm]
\hat{\pi}(\vecx) &= \frac{-i}{\sqrt{2V}} \, \sum_{\veck \in \mcalR} \sqrt{\o_{k}} \, \Big( \ha^{\phantom{\dagger}}_{\veck} \, e^{i \veck \cdot \vecx} - \ha^{\dagger}_{\veck} \, e^{- i \veck \cdot \vecx} \Big) \label{quantised pi} \quad ,
\end{align}
where $\o_{k}^{\phantom{1}} = \sqrt{m^{2} + k^{2\,}}$ and $k = \norm{\veck}\,$. The sum over $\mcalR$ is in practice a sum over $\mbbZ^{3}$, since the wave vectors are of the form $\veck = \mathbf{n} \,\, 2 \pi/L$ with $\mathbf{n} \in \mbbZ^{3}$. Imposing the f\mbox{}ield and momentum operators to satisfy the ``continuum'' CCR
\begin{align} \label{CCR phi pi}
\Big[ \hat{\phi}(\vecx) \, , \hat{\phi}(\vecy) \Big] &= \Big[ \hat{\pi}(\vecx) \, , \hat{\pi}(\vecy) \Big] = 0 \quad  , & \Big[ \hat{\phi}(\vecx) \, , \hat{\pi}(\vecy) \Big] &= i \, \d(\vecx - \vecy) \quad ,
\end{align}
amounts to postulating the following commutation relations for the operators $\ha$ and $\ha^{\dagger}$
\begin{align} \label{CR a adag}
\Big[ \, \ha^{\phantom{\dagger}}_{\veck} \, , \ha^{\phantom{\dagger}}_{\veck^{\p}} \Big] &= \Big[ \, \ha^{\dagger}_{\veck} \, , \ha^{\dagger}_{\veck^{\p}} \Big] = 0 \quad  , & \Big[ \, \ha^{\phantom{\dagger}}_{\veck} \, , \ha^{\dagger}_{\veck^{\p}} \Big] &= \d_{\veck , \veck^{\p}} \quad .
\end{align}
Neglecting in $\hat{H}_{_{0}}$ the (inf\mbox{}inite) ground state energy of the oscillators we obtain
\begin{align}
\hat{H}_{_{0}} &= \sum_{\veck \in \mcalR} \, \o_{k}^{\phantom{1}} \, \ha^{\dagger}_{\veck} \, \ha^{\phantom{\dagger}}_{\veck} \nn \\[1mm]
\hat{H}_{\textup{i}} &= \sum_{i = 1}^{2} \, \frac{q_{i}^{\phantom{\ast}}}{\sqrt{2V}} \, \sum_{\veck \in \mcalR} \frac{1}{\sqrt{\o_{k}}} \, \Big[ \, \ha^{\phantom{\dagger}}_{\veck} \,\, \ti{\r}^{\ast}_{i}(\veck) \, e^{i \veck \cdot \vecy_{\!i}} + \ha^{\dagger}_{\veck} \,\, \ti{\r}^{\phantom{\ast}}_{i}(\veck) \, e^{- i \veck \cdot \vecy_{\!i}} \Big] \quad , \label{HI HB a a*}
\end{align}
where $\ti{\r}^{\phantom{\ast}}_{i}$ is the Fourier transform of $\r_{i}$
\begin{equation} \label{rho Fourier transform}
\ti{\r}^{\phantom{\ast}}_{i}(\veck) = \int_{\mscrP} \! \r_{i}(\vecx) \, e^{-i \veck \cdot \vecx} \, d^{3}x \quad .
\end{equation}
Note that $\ti{\r}^{\ast}_{i}(\veck) = \ti{\r}^{\phantom{\ast}}_{i}(-\veck)$ as a consequence of $\r^{\phantom{\ast}}_{i}$ being real, and that if we take $\r^{\phantom{\ast}}_{i}$ to be spherically symmetric then $\ti{\r}^{\phantom{\ast}}_{i}$ depends only on the modulus $k\,$.

Introducing the operators $\hat{q}_{\veck}^{\phantom{\dg}}$ and $\hat{p}_{\veck}^{\phantom{\dg}}$ as follows
\begin{align} \label{qk pk}
\hat{q}_{\veck}^{\phantom{\dg}} &= \frac{1}{\sqrt{2}} \, \Big( \ha^{\phantom{\dg}}_{\veck} +  \ha^{\dg}_{\veck} \Big) & \hat{p}_{\veck}^{\phantom{\dg}} &= \frac{i}{\sqrt{2}} \, \Big( \ha^{\dagger}_{\veck} - \ha^{\phantom{\dg}}_{\veck} \Big) \quad ,
\end{align}
which therefore obey the CCR
\begin{align} \label{CCR q p}
\Big[ \, \hat{q}_{\veck}^{\phantom{\dg}} \, , \hat{q}_{\veck^{\p}}^{\phantom{\dg}} \Big] &= \Big[ \, \hat{p}_{\veck}^{\phantom{\dg}} \, , \hat{p}_{\veck^{\p}}^{\phantom{\dg}} \Big] = 0 \quad  , & \Big[ \, \hat{q}_{\veck}^{\phantom{\dg}} \, , \hat{p}_{\veck^{\p}}^{\phantom{\dg}} \Big] &= i \, \d_{\veck , \veck^{\p}} \quad ,
\end{align}
we obtain the following expression for the total Hamiltonian
\begin{equation} \label{total Hamiltonian}
\hat{H} = B + C + \frac{1}{2} \, \sum_{\veck \in \mcalR} \, \o_{k}^{\phantom{1}} \, \bigg[ \, \Big( \hat{p}_{\veck}^{\phantom{\dg}} + \s_{\!\veck}^{\phantom{\dg}} \Big)^{\! 2} + \Big( \hat{q}_{\veck}^{\phantom{\dg}} + \t_{\veck}^{\phantom{\dg}} \Big)^{\! 2} - 1 \, \bigg] \quad ,
\end{equation}
where
\begin{align} \label{tau e sigma}
\t_{\veck}^{\phantom{\dg}} &= \frac{g}{\sqrt{V \o_{k}^{3}}} \, \sum_{i = 1}^{2} \, q^{\phantom{\ast}}_{i} \, C^{\, i}_{\veck} & \s_{\!\veck}^{\phantom{\dg}} &= - \frac{g}{\sqrt{V \o_{k}^{3}}} \, \sum_{i = 1}^{2} \, q^{\phantom{\ast}}_{i} \, S^{\, i}_{\veck} \quad ,
\end{align}
with\footnote{$\Re[z]$ and $\Im[z]$ indicate respectively the real and the imaginary part of $z \in \mathbb{C}\,$.}
\begin{align}
C^{\, i}_{\veck} &= \Re \Big[ \ti{\r}^{\phantom{\ast}}_{i}(\veck) \Big] \, \cos \big( \veck \cdot \vecy_{\!i} \big) + \Im \Big[ \ti{\r}^{\phantom{\ast}}_{i}(\veck) \Big] \, \sin \big( \veck \cdot \vecy_{\!i} \big) \label{C_ik} \quad , \\[2mm]
S^{\, i}_{\veck} &= - \Im \Big[ \ti{\r}^{\phantom{\ast}}_{i}(\veck) \Big] \, \cos \big( \veck \cdot \vecy_{\!i} \big) + \Re \Big[ \ti{\r}^{\phantom{\ast}}_{i}(\veck) \Big] \, \sin \big( \veck \cdot \vecy_{\!i} \big) \quad . \label{S_ik}
\end{align}
The numbers $B$ and $C$ satisfy
\begin{equation} \label{B + C}
B + C = - \frac{1}{2} \, \sum_{\veck \in \mcalR} \, \o_{k}^{\phantom{1}} \, \Big( \s^{2}_{\veck} + \t^{2}_{\veck} \Big) \quad ,
\end{equation}
and indicating $g^{\phantom{\ast}}_{i} = g \, q^{\phantom{\ast}}_{i}$ they explicitly read
\begin{align}
\begin{split}
B &= - \frac{g_{_{1}} g_{_{2}}}{V} \, \sum_{\veck \in \mcalR} \, \frac{1}{\o_{k}^{2}} \, \bigg\{ \Re \Big[ \ti{\r}^{\phantom{\ast}}_{_{\!1}}\!(\veck) \, \ti{\r}^{\ast}_{_{2}}(\veck) \Big] \, \cos \Big[ \veck \cdot \big( \vecy_{_{\!1\!}} - \vecy_{_{\!2\!}} \big) \Big] + \\
&\hspace{45mm}+ \Im \Big[ \ti{\r}^{\phantom{\ast}}_{_{\!1}}\!(\veck) \, \ti{\r}^{\ast}_{_{2}}(\veck) \Big] \, \sin \Big[ \veck \cdot \big( \vecy_{_{\!1\!}} - \vecy_{_{\!2\!}} \big) \Big] \bigg\} \quad ,
\end{split} \label{B} \\[2mm]
C &= - \sum_{i = 1}^{2} \,\, \frac{g_{i}^{2}}{2 V} \sum_{\veck \in \mcalR} \, \frac{1}{\o_{k}^{2}} \, \babs{\ti{\r}^{\phantom{\ast}}_{i}(\veck)}^{2} \label{C} \quad .
\end{align}

\subsection{The physical meaning of $B$ and $C$}
\label{subsec: The numbers B and C}

Comparing these expressions with the corresponding ones in \cite{Van Hove 1952,Sbisa 2020}, it is apparent that the expressions (\ref{total Hamiltonian}) and (\ref{B + C}) for the Hamiltonian have exactly the same form as those for point-like sources. The dif\mbox{}ference shows only in the explicit expression of $\t_{\veck}^{\phantom{\dg}}$ and $\s_{\veck}^{\phantom{\dg}}$ in terms of the parameters of the model.

Regarding $B$ and $C$, their expressions in the point-like case are
\begin{align}
B_{\textup{pt}} &= - \frac{g_{_{1}} g_{_{2}}}{V} \, \sum_{\veck \in \mcalR} \, \frac{1}{\o_{k}^{2}} \, \cos \Big[ \veck \cdot \big( \vecy_{_{\!1\!}} - \vecy_{_{\!2\!}} \big) \Big] \quad , \label{B sharp} \\[2mm]
C_{\textup{pt}} &= - \frac{g^{2}_{_{1}} + g^{2}_{_{2}}}{2 V} \, \sum_{\veck \in \mcalR} \, \frac{1}{\o_{k}^{2}} \label{C sharp} \quad ,
\end{align}
and, comparing with (\ref{B}) and (\ref{C}), it is apparent that giving a f\mbox{}inite spatial width to the sources introduces a modulation in the sum over the reciprocal lattice. This modulation depends only on the modulus $k$ when $\r_{_{1}}\!$ and $\r_{_{2}}\!$ are spherically symmetric. The expressions (\ref{B sharp}) and (\ref{C sharp}) are reproduced by setting $\ti{\r}^{\phantom{\ast}}_{_{1}}\! = \ti{\r}^{\phantom{\ast}}_{_{2}}\! = 1$ in (\ref{B}) and (\ref{C}), which is compatible with the idea that ``the Fourier transform of a Dirac delta is a constant''.

The central observation is that, as we discuss in section \ref{app subsec convergence} of the appendix, when $\r_{_{1}}\!$ and $\r_{_{2}}\!$ have compact support the modulations due to $\ti{\r}^{\phantom{\ast}}_{_{1}}\!$ and $\ti{\r}^{\phantom{\ast}}_{_{2}}\!$ in the expressions (\ref{B}) and (\ref{C}) are rapidly decreasing, and so the inf\mbox{}inite sums in (\ref{B}) and (\ref{C}) are absolutely convergent. In comparison, in the point-like sources case the inf\mbox{}inite sum in (\ref{C sharp}) is divergent, and the sum in (\ref{B sharp}) has to be understood in a suitable way to be considered convergent. The model with spatially distributed sources is therefore non-singular, unlike the point-like description where $C$ diverges and it is necessary to arbitrarily discard it to progress with the analysis. The search for a less unrealistic description of the sources has automatically provided a model where the divergence is regularized.

Let us look more closely at the meaning of $C\,$, which is somehow a mysterious object in \cite{Van Hove 1952}. This is best seen in the inf\mbox{}inite volume limit, which is well def\mbox{}ined as we prove in section \ref{subsubsec: Existence} of the appendix. In section \ref{subsubsec: Potential-mediated} we show that in this limit we have
\begin{align}
B &\xrightarrow[V \to \infty]{} - g_{_{1}} g_{_{2}} \iint_{\mbbR^{6}} r_{_{\!1}}(\vecx - \vecy_{_{\!1\!}}) \, r_{_{\!2}}(\vecz - \vecy_{_{\!2\!}}) \,\, \frac{1}{4 \pi} \, \frac{e^{-m \norm{\vecx - \vecz}}}{\norm{\vecx - \vecz}} \,\, d^{3}x \, d^{3}z \quad , \label{B inf vol} \\[2mm] 
C &\xrightarrow[V \to \infty]{} - \frac{1}{2} \, \sum_{i = 1}^{2} \,\, g^{2}_{i} \iint_{\mbbR^{6}} r_{i}(\vecx) \, r_{i}(\vecz) \,\, \frac{1}{4 \pi} \, \frac{e^{- m \norm{\vecx - \vecz}}}{\norm{\vecx - \vecz}} \,\, d^{3}x \, d^{3}z \quad , \label{C inf vol}
\end{align}
where the auxiliary functions $r_{_{\!1}}\!$ and $r_{_{\!2}}\!$ are def\mbox{}ined as in (\ref{app: auxiliary function}) and are closely related to $\r_{_{1}}\!$ and $\r_{_{2}}$. It is apparent that $C$ is nothing else that the total self-interaction energy of the sources ($B$ is the sum of the pair-wise interaction energy between dif\mbox{}ferent sources, as was already recognized in \cite{Van Hove 1952}). Considering distributed sources therefore permits to recognize that the numbers $B$ and $C$ are two aspects of the same physical mechanism, the interaction via the Yukawa potential mediated by the f\mbox{}ield. If $\r_{_{1}}\!$ and $\r_{_{2}}\!$ become more and more peaked, the number $C$ grows unboundedly while $B$ doesn't, since in the latter case we are coupling charge distributions which are spatially separated. This reveals that the divergence of $C_{\textup{pt}}$ is just a spurious artifact due to a too crude description of the internal structure of the sources, and permits to appreciate that treating $B$ and $C$ on dif\mbox{}ferent grounds, as done in \cite{Van Hove 1952}, is artif\mbox{}icial.

\subsection{Energy eigenstates and spectrum}
\label{subsec: eigenstates and spectrum}

We now turn to the derivation of the exact solutions of the model. In \cite{Van Hove 1952} this is achieved by representing the abstract canonical operators $\hat{q}_{\veck}^{\phantom{\dg}}$ and $\hat{p}_{\veck}^{\phantom{\dg}}$, and therefore the Hamiltonian, as explicit operators on a space of functionals. We follow the same path here. More precisely, we choose a representation of the CCR in which the operators $\hat{q}_{\veck}^{\phantom{\dg}}$ are diagonal. The state vectors of the f\mbox{}ield are represented as functionals $\Phi(\{q\})$, where $\{q\} = \{ q_{\veck}^{\phantom{\dg}} \}_{\veck \in \mcalR}^{\phantom{\dg}}$ indicates a family of real numbers indexed by the vectors of the reciprocal lattice. The abstract operators $\hat{q}_{\veck}^{\phantom{\dg}}$ and $\hat{p}_{\veck}^{\phantom{\dg}}$ are realized respectively as the operator which multiply $\Phi(\{q\})$ by $q_{\veck}^{\phantom{\dg}}\,$, and as the derivative operator with respect to $q_{\veck}^{\phantom{\dg}}$ multiplied by the imaginary unit.

The (total) Hamiltonian operator is therefore realized as the dif\mbox{}ferential operator
\begin{equation} \label{mathcal H}
\hat{\mathcal{H}} \,\, \Phi = \Bigg\{ \, B \, + \, C \, + \frac{1}{2} \, \sum_{\veck \in \mcalR} \, \o_{k}^{\phantom{1}} \, \Bigg[ \bigg( \! - i \frac{\de}{\de q_{\veck}^{\phantom{\dg}}} + \s_{\!\veck}^{\phantom{\dg}} \bigg)^{\!\! 2} + \big( q_{\veck}^{\phantom{\dg}} + \t_{\veck}^{\phantom{\dg}} \big)^{2} - 1 \, \Bigg] \Bigg\} \, \Phi \,\,\, ,
\end{equation}
whose eigenfunctionals are the inf\mbox{}inite products
\begin{equation} \label{eigenfunctions}
\Phi_{\{ n \}}  \big( \{ q \} \big) = \prod_{\veck \in \mcalR} \vf_{\!n(\veck)}^{\veck} \big( q_{\veck}^{\phantom{\dg}} \big) \quad ,
\end{equation}
where the functions $\vf_{n}^{\veck}$ are def\mbox{}ined by
\begin{equation} \label{varphi}
\vf_{n}^{\veck} \big( q_{\veck}^{\phantom{\dg}} \big) = e^{-i \, \s_{\!\veck}^{\phantom{\dg}} q_{\veck}^{\phantom{\dg}}} \,\, \ps_{n} \Big( q_{\veck}^{\phantom{\dg}} + \t_{\veck}^{\phantom{\dg}} \Big) \quad ,
\end{equation}
and $\ps_{n}$ is the standard Hermite function of $n$-th degree
\begin{equation}
\ps_{n}\big( x \big) = \frac{(-1)^{n}}{\sqrt{2^{n} \, n! \, \sqrt{\pi}}} \,\, e^{\frac{x^{2}}{2}} \frac{d^{n}}{dx^{n}} e^{-x^{2}} \quad .
\end{equation}
The energy eigenvalue associated to $\Phi_{\{ n \}}$ is
\begin{equation}
\label{spectrum}
E_{\{ n \}} = B + C + \sum_{\veck \in \mcalR} \, n(\veck) \,\, \o_{k}^{\phantom{1}} \quad ,
\end{equation}
so the energy spectrum is basically the same as that of a free relativistic neutral scalar f\mbox{}ield of mass $m\,$, with the only dif\mbox{}ference of being rigidly shifted by the Yukawa-like interaction energy $B + C$ of the f\mbox{}ixed charge distributions.

Formally, the only change in the expressions (\ref{mathcal H})--(\ref{spectrum}) with respect to the corresponding equations in \cite{Van Hove 1952,Sbisa 2020} is that $C$ is now included in $\hat{\mathcal{H}}$ and therefore in its spectrum, and not discarded anymore. Apart from this, the inf\mbox{}luence of $\r_{_{1}\!}$ and $\r_{_{2}\!}$ is felt only through the phase and shift factors $\s_{\!\veck}^{\phantom{\dg}}$ and $\t_{\veck}^{\phantom{\dg}}\,$.

\section{The orthogonality of the spaces of state vectors}
\label{sec: Orthogonality spaces state vectors}

From the analysis of the previous section it follows that each family of non-negative integers $\{ n \} = \{n(\veck)\}_{\veck \in \mcalR}$ individuates an energy eigenstate $\Phi_{\{ n \}}\,$. To def\mbox{}ine the space of state vectors it is then necessary to decide which class of families is associated to physically acceptable states. In \cite{Van Hove 1952}, this problem is settled by noting that the only energy eigenstates which can be excited from the ground state using a f\mbox{}inite amount of energy are those whose associated sequence possesses a f\mbox{}inite number of non-zero elements. Therefore, the space of state vectors is def\mbox{}ined as the Hilbert space generated by the stationary states of f\mbox{}inite energy, that is by the $\Phi_{\{ n \}}$ for which the non-zero $n(\veck)$ are in a f\mbox{}inite number. This reasoning holds equally well in our case, so we adopt the same def\mbox{}inition of space of state vectors. It follows that, as in \cite{Van Hove 1952}, the latter has the structure of non-separable Hilbert space of the inf\mbox{}inite direct product type, as introduced by von Neumann \cite{von Neumann Comp Math 1939}.

\subsection{Disappearance of the orthogonality}

A central result of \cite{Van Hove 1952} is that the spaces of state vectors relative to dif\mbox{}ferent parameter choices are orthogonal. 
We now want to investigate whether this result continues to hold in the model with spatially distributed sources. Before discussing the details, it is worthwhile to clarify that, since the model is invariant with respect to the simultaneous rescaling $g \to c \, g\,$, $q_{_{1}} \to q_{_{1}}/c$ and $q_{_{2}} \to q_{_{2}}/c$ by a real number $c\,$, it is necessary to work here with the products $g_{_{1\!}} = g \, q_{_{1\!}}$ and $g_{_{2\!}} = g \, q_{_{2\!}}$ to avoid this degeneracy undermining our conclusions. So the parameters of the model are: the product of the coupling constant and the total charges $g_{_{1}}$, $g_{_{2}}$; the position of the charges $\vecy_{_{\!1}} , \vecy_{_{\!2}}$; the distributions $\r_{_{1}} , \r_{_{2}}$.

Be then $\G = \big\{ (g_{_{1}}, \vecy_{_{\!1}}, \r_{_{1\!}})\, , (g_{_{2}}, \vecy_{_{\!2}}, \r_{_{2\!}}) \big\}$ and $\bar{\Gamma} = \big\{ (\bar{g}_{_{1}}, \vecz_{_{1}}, \z_{_{1\!}}) \, , (\bar{g}_{_{2}} , \vecz_{_{2}}, \z_{_{2\!}}) \big\}$ two choices for the parameters of the model, with $\vecy_{_{\!1}}\! \neq \vecy_{_{\!2}}$, $\vecz_{_{1}}\! \neq \vecz_{_{2}}\!$ and $\r_{_{1}}, \r_{_{2}}, \z_{_{1}}, \z_{_{2}}\!$ all compactly supported. As shown in \cite{Van Hove 1952,Sbisa 2020}, the spaces of state vectors relative to dif\mbox{}ferent parameters choices are orthogonal if and only if the corresponding ground states are orthogonal. Let us therefore assume $\G \neq \bar{\Gamma}\,$, and indicate with $\Phi_{_{\!\{ 0 \}}}\!$ the ground state when the parameters choice is $\G$, while $\bar{\Phi}_{_{\!\{ 0 \}}}\!$ indicates the ground state when the parameters choice is $\bar{\Gamma}$. The calculation of the scalar product between $\bar{\Phi}_{_{\!\{ 0 \}}}\!$ and $\Phi_{_{\!\{ 0 \}}}\!$ is formally the same as in \cite{Van Hove 1952,Sbisa 2020}, so we get
\begin{equation*}
\Babs{\Big\langle \bar{\Phi}_{_{\!\{ 0 \}\!}} \, , \Phi_{_{\!\{ 0 \}\!}} \Big\rangle} = \prod_{\veck \in \mcalR} \Babs{\Big\langle \bar{\vf}_{_{\!0\phantom{(\veck)}}}^{\veck} \!\!\! , \vf_{_{\!0\phantom{(\veck)}}}^{\veck} \!\!\!\! \Big\rangle} = \exp \Bigg\{ - \frac{1}{4} \sum_{\veck \in \mcalR} \bigg[ \Big( \s_{\veck}^{\phantom{\dg}} - \, \bar{\sigma}_{\veck}^{\phantom{\dg}} \Big)^{\!2} + \Big( \t_{\veck}^{\phantom{\dg}} - \, \bar{\tau}_{\veck}^{\phantom{\dg}} \Big)^{\!2} \, \bigg] \Bigg\} \quad ,
\end{equation*}
and inserting the expressions (\ref{tau e sigma})--(\ref{S_ik}) for the phase and shift factors we now f\mbox{}ind
\begin{multline}
\Big( \s_{\veck}^{\phantom{\dg}} - \bar{\sigma}_{\veck}^{\phantom{\dg}} \Big)^{\!2} + \Big( \t_{\veck}^{\phantom{\dg}} - \bar{\tau}_{\veck}^{\phantom{\dg}} \Big)^{\!2} \leq \frac{\sqrt{2}}{V \o_{k}^{3}} \, \bigg\{ \Big[ \, \babs{g_{_{1\,}} \ti{\r}^{\phantom{\ast}}_{_{1}}\!(\veck)} + \babs{g_{_{2\,}} \ti{\r}^{\phantom{\ast}}_{_{2}}\!(\veck)} \, \Big]^{2} + \\
+ 2 \, \sum_{i,j = 1}^{2} \babs{g_{i} \, \ti{\r}^{\phantom{\ast}}_{i}\!(\veck)} \, \babs{\bar{g}_{j} \, \ti{\z}^{\phantom{\ast}}_{j}\!(\veck)} + \Big[ \, \babs{\bar{g}_{_{1\,}} \ti{\z}^{\phantom{\ast}}_{_{1}}\!(\veck)} + \babs{\bar{g}_{_{2\,}} \ti{\z}^{\phantom{\ast}}_{_{2}}\!(\veck)} \, \Big]^{2} \bigg\} \quad .
\end{multline}
Since $\r_{_{1}}, \r_{_{2}}, \z_{_{1}}\!$ and $\z_{_{2}}\!$ are all compactly supported, we can use exactly the same arguments of section \ref{app subsec convergence} of the appendix to infer that the inf\mbox{}inite sum
\begin{multline}
\frac{\sqrt{2}}{V} \, \sum_{\veck \in \mcalR} \frac{1}{\o_{k}^{3}} \, \bigg\{ \Big[ \, \babs{g_{_{1\,}} \ti{\r}^{\phantom{\ast}}_{_{1}}\!(\veck)} + \babs{g_{_{2\,}} \ti{\r}^{\phantom{\ast}}_{_{2}}\!(\veck)} \, \Big]^{2} + \\
+ 2 \, \sum_{i,j = 1}^{2} \babs{g_{i} \, \ti{\r}^{\phantom{\ast}}_{i}\!(\veck)} \, \babs{\bar{g}_{j} \, \ti{\z}^{\phantom{\ast}}_{j}\!(\veck)} + \Big[ \, \babs{\bar{g}_{_{1\,}} \ti{\z}^{\phantom{\ast}}_{_{1}}\!(\veck)} + \babs{\bar{g}_{_{2\,}} \ti{\z}^{\phantom{\ast}}_{_{2}}\!(\veck)} \, \Big]^{2} \bigg\}
\end{multline}
converges (here we even have $\o^{3}$ instead of $\o^{2}$ in the denominator). It follows that $\sum_{\veck \in \mcalR} \Big[ \big( \s_{\!\veck}^{\phantom{\dg}} - \bar{\sigma}_{\!\veck}^{\phantom{\dg}} \big)^{\!2} + \big( \t_{\veck}^{\phantom{\dg}} - \bar{\tau}_{\veck}^{\phantom{\dg}} \big)^{\!2} \, \Big]$ converges (absolutely), which in turn implies that 
\begin{equation} \label{scalar product 0 neq 0}
\Babs{\Big\langle \bar{\Phi}_{_{\!\{ 0 \}\!}} \, , \Phi_{_{\!\{ 0 \}\!}} \Big\rangle} \neq 0 \quad .
\end{equation}
Therefore we conclude that, when the sources are spatially distributed, the spaces of state vectors relative to dif\mbox{}ferent parameter choices are \emph{never} orthogonal, no matter how peaked the distributions. This result is compatible with the exposition of \cite{Reed Simon III} regarding a relativistic and massive scalar quantum f\mbox{}ield interacting with an external potential, although in the latter case the source is assumed to be of compact support also in the time variable.

\subsection{Divergence \emph{vs} orthogonality}

To be more specif\mbox{}ic, let us focus on the case where $\bar{g}_{_{1}}\! = \bar{g}_{_{2}}\! = 0\,$, that is comparing the energy eigenstates associated to the parameters choice $\G = \big\{ (g_{_{1}}, \vecy_{_{\!1}}, \r_{_{1\!}} ) \, , (g_{_{2}}, \vecy_{_{\!2}}, \r_{_{2\!}} ) \big\}$ with the energy eigenstates of the \emph{free} Hamiltonian. In this case we have $\bar{\tau}_{\veck}^{\phantom{\dg}} = \bar{\sigma}_{\!\veck}^{\phantom{\dg}} = 0$ so from the previous section we get
\begin{equation} \label{scalar product free int}
\Babs{\Big\langle \Phi_{_{\!\{ 0 \}\!}}^{\textup{free}} \, , \Phi_{_{\!\{ 0 \}\!}} \Big\rangle} = \exp \Bigg[ \! - \frac{1}{4} \sum_{\veck \in \mcalR} \Big( \s_{\veck}^{2} + \t_{\veck}^{2} \Big) \Bigg] \quad ,
\end{equation}
which implies that the space of state vectors labeled by $\G$ and the space of free state vectors are orthogonal if and only if 
\begin{equation} \label{orthogonality condition}
\sum_{\veck \in \mcalR} \Big( \s_{\veck}^{2} + \t_{\veck}^{2} \Big) = \infty \quad .
\end{equation}
On the other hand, recalling from (\ref{B + C}) that
\begin{equation} \label{B + C again}
B + C = - \frac{1}{2} \, \sum_{\veck \in \mcalR} \, \o_{k}^{\phantom{1}} \, \Big( \s^{2}_{\veck} + \t^{2}_{\veck} \Big)
\end{equation}
and comparing the latter with (\ref{orthogonality condition}), we note that the condition of the ground state energy $B + C$ being divergent is closely related to the condition of orthogonality of the free and interacting spaces of state vectors, being just slightly weaker (due to the factor $\o_{k}^{\phantom{1}}$).

This implies that it is not possible to achieve the phenomenon of orthogonality of the free and interacting spaces of state vectors with a well-def\mbox{}ined model, that is one where $B + C$ is f\mbox{}inite. This follows from the fact that if (\ref{orthogonality condition}) holds, then surely (\ref{B + C again}) diverges. On the other hand, it may be possible to have a model which is singular and yet where the free and interacting spaces of state vectors are not orthogonal. For the sake of completeness, let us look for simple conditions for this (admittedly not very interesting) possibility to happen. Let us assume that the two functions $\r_{_{1}}\!$ and $\r_{_{2}}\!$ are equal ($ = \r$) and spherically symmetric, and work in the inf\mbox{}inite volume limit. Recalling from (\ref{tau e sigma})--(\ref{S_ik}) that in this case
\begin{equation}
\s_{\veck}^{2} + \t_{\veck}^{2} = \frac{\babs{\ti{\r}(k)}^{2}}{V \o_{k}^{3}} \, \bigg\{ g_{_{1}}^{2} + g_{_{2}}^{2} + 2 \, g_{_{1}} g_{_{2}} \cos \Big[ \veck \cdot \big(\vecy_{_{\!1}}\! - \vecy_{_{\!2\!}}\big) \Big] \bigg\} \quad ,
\end{equation}
any $\ti{\r}$ such that
\begin{equation}
0 \neq \lim_{k \to \infty} k^{\b} \, \babs{\ti{\r}(k)}^{2} < \infty
\end{equation}
for some $\b$ in the interval $0 < \b \leq 1$ satisf\mbox{}ies the condition of $B + C$ diverging while the inf\mbox{}inite sum in (\ref{orthogonality condition}) being f\mbox{}inite. In this case $\ti{\r}$ is not even square-integrable. In particular, the same results we obtained for compactly supported distributions would hold also if, in the inf\mbox{}inite volume limit, we considered $\r_{_{1}}\!$ and $\r_{_{2}}\!$ to be rapidly decreasing (for example taking them equal to the electronic distribution of the ground state of the hydrogen atom).

\subsubsection{Comparison with the point-like case}

It is interesting to see explicitly how the orthogonality result of \cite{Van Hove 1952} is recovered in our description. Going back to the general case $\bar{g}_{_{1}}\! \neq 0\,$, $\bar{g}_{_{2}}\! \neq 0\,$, let us consider for simplicity the case where the distributions are all equal, that is $\r^{\phantom{\ast}}_{_{1}}\! = \r^{\phantom{\ast}}_{_{2}}\! = \z^{\phantom{\ast}}_{_{1}}\! = \z^{\phantom{\ast}}_{_{2}}\! = \r\,$. We obtain
\begin{multline}
\sum_{\veck \in \mcalR} \Big[ \big( \s_{\veck}^{\phantom{\dg}} - \bar{\sigma}_{\veck}^{\phantom{\dg}} \big)^{\!2} + \big( \t_{\veck}^{\phantom{\dg}} - \bar{\tau}_{\veck}^{\phantom{\dg}} \big)^{\!2} \, \Big] = \\
= \frac{1}{V} \sum_{\veck \in \mcalR} \, \frac{\babs{\ti{\r}(\veck)}^{2}}{\o_{k}^{3}} \, \bigg\{ g_{_{1}}^{2} + g_{_{2}}^{2} + \bar{g}_{_{1}}^{2} + \bar{g}_{_{2}}^{2} + 2 \, g_{_{1}} g_{_{2}} \cos \Big[ \veck \cdot \big(\vecy_{_{\!1}}\! - \vecy_{_{\!2\!}}\big) \Big] + \\
+ 2 \, \bar{g}_{_{1}} \bar{g}_{_{2}} \cos \Big[ \veck \cdot \big(\vecz_{_{1\!}} - \vecz_{_{2\!}}\big) \Big] - 2 \sum_{i, j = 1}^{2} g_{_{i}} \bar{g}_{_{j}} \cos \Big[ \veck \cdot \big(\vecy_{i} - \vecz_{j}\big) \Big] \bigg\} \quad ,
\end{multline}
which again converges absolutely since $\ti{\r}$ is rapidly decreasing. The corresponding result of \cite{Van Hove 1952,Sbisa 2020} can be retrieved by setting $\babs{\ti{\r}(\veck)} = 1\,$. In this case the denominator $\o_{k}^{3}$ is not suf\mbox{}f\mbox{}icient by itself to make the inf\mbox{}inite sum converge, and for the convergence to take place suitable cancellations need to happen inside the round parenthesis in the right hand side (even in this case, though, the inf\mbox{}inite sum does not converges absolutely and we need to def\mbox{}ine the order of summation in a suitable way to be able to consider the sum convergent). The condition for such cancellations to happen turns out to be $\G = \bar{\Gamma}\,$, which is the celebrated result of \cite{Van Hove 1952}.

\subsubsection{Algebraic treatment}

It is worthwhile to remind that the orthogonality condition (\ref{orthogonality condition}) can be also derived by algebraic means. Following \cite{Wightman Schweber 1955}, we introduce the operators
\begin{align}
\hat{A}^{\phantom{\dg}}_{\veck} &= \frac{1}{\sqrt{2}} \, \bigg[ \, \hat{q}_{\veck}^{\phantom{\dg}} + \t_{\veck}^{\phantom{\dg}} + i \Big( \hat{p}_{\veck}^{\phantom{\dg}} + \s_{\!\veck}^{\phantom{\dg}} \Big) \bigg] = \ha^{\phantom{\dg}}_{\veck} + \a^{\phantom{\dg}}_{\veck} \\[2mm]
\hat{A}^{\dg}_{\veck} &= \frac{1}{\sqrt{2}} \, \bigg[ \, \hat{q}_{\veck}^{\phantom{\dg}} + \t_{\veck}^{\phantom{\dg}} - i \Big( \hat{p}_{\veck}^{\phantom{\dg}} + \s_{\!\veck}^{\phantom{\dg}} \Big) \bigg] = \ha^{\dg}_{\veck} + \a^{\ast}_{\veck} \quad ,
\end{align}
where $\a_{\veck}^{\phantom{\dg}} = \t_{\veck}^{\phantom{\dg}} + i \, \s_{\!\veck}^{\phantom{\dg}}\,$. Since the commutation rules for $\hat{A}^{\phantom{\dg}}_{\veck}$ and $\hat{A}^{\dg}_{\veck}$ are equal to those for $\ha^{\phantom{\dg}}_{\veck}$ and $\ha^{\dg}_{\veck}\,$, to wit (\ref{CR a adag}), the total (interacting) Hamiltonian can be written as
\begin{equation} \label{algebraic interacting Hamiltonian}
\hat{H} = B + C + \sum_{\veck \in \mcalR} \o_{k}^{\phantom{1}} \,\, \hat{A}^{\dg}_{\veck} \, \hat{A}^{\phantom{\dg}}_{\veck} \quad ,
\end{equation}
so its structure is formally similar to that of the free Hamiltonian.

Coherently with the discussion of section \ref{subsec: eigenstates and spectrum}, the operators $\hat{A}$ and $\hat{A}^{\dg}$ are def\mbox{}ined on a non-separable Hilbert space $\mscrH$ of the inf\mbox{}inite direct product type. Let us realize the non-constant part of the Hamiltonian $\hat{H}$ as the limit of a sequence of operators indexed by $n \in \mbbN\,$, such that the $n$-th element of the sequence reads
\begin{equation}
\sum_{\veck \in \mcalR , \, \norm{\veck} \leq n} \o_{k}^{\phantom{1}} \,\, \hat{A}^{\dg}_{\veck} \, \hat{A}^{\phantom{\dg}}_{\veck} \quad .
\end{equation}
Although the sequence of operators is def\mbox{}ined on (a dense subset of) $\mscrH$, the limit as $n \to \infty$ of this sequence converges only on a (dense subset of a) \emph{separable subspace} $\mcalH_{_{\G}} \subset \mscrH\,$, which in general depends on the choice of the parameters collectively indicated with $\G$ \cite{Wightman Schweber 1955}. Wightman and Schweber assert that, if
\begin{equation}
\sum_{\veck \in \mcalR} \babs{\a_{\veck}^{\phantom{\ast}}}^{2} < \infty \quad ,
\end{equation}
then all the subspaces $\mcalH_{_{\G}}$ coincide, that is $\mcalH_{_{\G}} = \mcalH_{_{\bar{\Gamma}}}$ even if $\G \neq \bar{\Gamma}$. On the other hand, if
\begin{equation}
\sum_{\veck \in \mcalR} \babs{\a_{\veck}^{\phantom{\ast}}}^{2} = \infty \quad ,
\end{equation}
then the subspaces $\mcalH_{_{\G}}$ corresponding to dif\mbox{}ferent parameters choices are orthogonal, that is $\G \neq \bar{\Gamma}$ implies $\mcalH_{_{\G}} \perp \mcalH_{_{\bar{\Gamma}}}\,$. The connection with the previous analysis, and in particular with the condition (\ref{orthogonality condition}), becomes evident once we recognize that
\begin{equation}
\babs{\a_{\veck}^{\phantom{\ast}}}^{2} = \s_{\veck}^{2} + \t_{\veck}^{2} \quad ,
\end{equation}
and that (\ref{scalar product free int}) can be expressed as
\begin{equation*}
\Babs{\Big\langle \Phi_{_{\!\{ 0 \}\!}}^{\textup{free}} \, , \Phi_{_{\!\{ 0 \}\!}} \Big\rangle} = \exp \Bigg( \!\! - \frac{1}{4} \sum_{\veck \in \mcalR} \babs{\a_{\veck}^{\phantom{\ast}}}^{2} \Bigg) \quad .
\end{equation*}

\section{Discussion}
\label{sec: Discussion}

The results of the previous sections unambiguously show that the divergence of the ground state energy $B + C$ of the model is strongly correlated with the property of orthogonality of the spaces of state vectors. The common underlying cause of both is the non convergence of the inf\mbox{}inite sum $\sum_{\veck \in \mcalR} \big( \s_{\!\veck}^{2} + \t_{\veck}^{2} \big)\,$. The two conditions are strictly speaking not equivalent, but the important point for our purposes is that it is impossible to have orthogonality while having a f\mbox{}inite ground state energy.

This prompts a revision of the validity of Van Hove's results. It is useful in this sense to identify two logically distinct parts in Van Hove's article. The f\mbox{}irst is the derivation, from the toy model of a scalar f\mbox{}ield interacting linearly with point-like sources, of the quantum Hamiltonian
\begin{subequations}
\begin{align}
\hat{H}_{\textup{pt}} &= B + \frac{1}{2} \, \sum_{\veck \in \mcalR} \, \o_{k}^{\phantom{1}} \, \bigg[ \, \Big( \hat{p}_{\veck}^{\phantom{\dg}} + \s_{\!\veck}^{\phantom{\dg}} \Big)^{\! 2} + \Big( \hat{q}_{\veck}^{\phantom{\dg}} + \t_{\veck}^{\phantom{\dg}} \Big)^{\! 2} - 1 \, \bigg] \quad , \label{Fluo 1} \\[2mm]
\sum_{\veck \in \mcalR} \, &\Big( \s_{\veck}^{2} + \t_{\veck}^{2} \Big) = \infty \quad , \label{Fluo 2}
\end{align}
\end{subequations}
the condition (\ref{Fluo 2}) being a def\mbox{}ining property of the Hamiltonian. Note that $C$ is discarded, while $B$, which is f\mbox{}inite, is kept. The second part is the study of this Hamiltonian and its phenomenology. The validity of both parts contribute to the relevance of the article as a whole. We discuss below the consequence of our results for the validity of each of these two parts. To avoid cumbersome phrases, below we use the term ``Van Hove's model'' to indicate the model of a scalar f\mbox{}ield interacting linearly with point-like, inf\mbox{}initely massive sources, without a priori identifying it with the Hamiltonian (\ref{Fluo 1})--(\ref{Fluo 2}).

\subsection{The model and the quantum Hamiltonian}

\subsubsection{On neglecting inf\mbox{}inities}

One possible justif\mbox{}ication for associating the quantum Hamiltonian (\ref{Fluo 1})--(\ref{Fluo 2}) to Van Hove's model is the assertion that it is a common practice to neglect inf\mbox{}inities in QFT. It could in fact be argued that, since the ground state energy of the oscillators is neglected right after quantizing, it should be safe to neglect any other inf\mbox{}inite contribution to the ground state energy of the system. From this point of view, the question seems to boil down to whether and when it is justif\mbox{}ied to ``neglect inf\mbox{}inities''. It cannot be avoided to note how the fact that neglecting inf\mbox{}inities has become routine in modern physics is playing here a confounding role.

The case of $C$ and that of the inf\mbox{}inite energy contribution coming from the ground state of the oscillators are, however, qualitatively dif\mbox{}ferent. The latter is of kinematical character: it is present whenever a scalar f\mbox{}ield is quantized and is independent of the dynamics of the specif\mbox{}ic model under consideration (i.e.\ from the interaction Hamiltonian). Specif\mbox{}ically, in our case the partial sums associated to the diverging series are independent of the value of the coupling constant and of the position of the charges. On the other hand, the contribution from $C$ has a dynamical character: the partial sums associated to the diverging series depend on the coupling constant and on the distribution of the charges. This is a signal that the divergence of $C$ may be pointing to a relevant dynamical mechanism which is not being taken into account appropriately.

We believe that the discussion in section \ref{subsec: The numbers B and C} makes a strong point: $B$ and $C$ are not dif\mbox{}ferent in essence, and both have an important \emph{physical} meaning. Therefore the divergence of $C$ is not to be seen as something which can be lightly discarded, on the contrary it signals that the original model, with point-like sources, is ill-def\mbox{}ined.

\subsubsection{On treating $B$ and $C$ dif\mbox{}ferently}

To exemplify how artif\mbox{}icial it is to treat $B$ and $C$ on dif\mbox{}ferent grounds, it is useful to reconsider the discussion in \cite{Van Hove 1952} (recalled in section 5.3 of \cite{Sbisa 2020}) about the applicability of the perturbative approach to evaluate the interaction energy $B$ between dif\mbox{}ferent sources, and the scalar product between the free and the interacting ground state.

In the f\mbox{}irst case, the second order perturbative correction to the interaction energy is evaluated imposing a momentum cut-of\mbox{}f, ``excluding every transition where one boson is emitted and absorbed by the same source'' (in Van Hove's words). The result indeed remains f\mbox{}inite when removing the cut-of\mbox{}f, causing Van Hove to conclude that the perturbative method is applicable to estimate $B\,$. However, the total second order perturbative correction to the ground state energy explicitly reads
\begin{equation}
E_{^{\{ 0\}}}^{_{(2)}}(\La) = - g^{2} \, \frac{q_{_{1}}^{2} + q_{_{2}}^{2}}{2 V} \sum_{\veck \in \mcalR_{_{\!\La}}} \! \frac{1}{\o_{k}^{2}} \, - g^{2} \, \frac{q_{_{1}} q_{_{2}}}{V} \sum_{\veck \in \mcalR_{_{\!\La}}} \! \frac{\cos \Big[ \veck \cdot \big( \vecy_{_{\!1\!}} - \vecy_{_{\!2\!}} \big) \Big]}{\o_{k}^{2}} \quad ,
\end{equation}
where $\La$ is the cut-of\mbox{}f and $\mcalR_{_{\!\La}}$ indicates the set of vectors of the reciprocal lattice of modulus $\leq \La\,$. Therefore, what Van Hove is really doing is to discard the addend multiplied by $q^{2}_{_{1}} + q^{2}_{_{2}}$, which after removing the cut-of\mbox{}f would give rise to the number $C\,$. This is the term which corresponds to ``bosons being emitted and absorbed by the same source'' (more precisely, the contribution associated to $q^{2}_{i}$ corresponds to ``bosons being emitted and absorbed by the source of charge $q_{i}$''). Only the addend multiplied by the mixed product $q_{_{1}} q_{_{2}}\!$ is kept, which after removing the cut-of\mbox{}f ($\La \to \infty$) gives rise to $B$ and remains f\mbox{}inite (if the summation is understood appropriately).

Consider now the second case, where the (absolute value of the) scalar product $a$ between the free and the interacting ground state is considered. Imposing again a momentum cut-of\mbox{}f $\La$ and working at the second order in perturbations, the result
\begin{equation}
a^{_{(2)\!}}(\La) =  1 - \frac{g^{2}}{4 V} \sum_{\veck \in \mcalR_{_{\!\La}}} \! \frac{1}{\o_{k}^{3}} \, \bigg\{ q_{_{1}}^{2} + q_{_{2}}^{2} + 2 \, q_{_{1}} q_{_{2}} \cos \Big[ \veck \cdot \big( \vecy_{_{\!1\!}} - \vecy_{_{\!2\!}} \big) \Big] \bigg\}
\end{equation}
displays again the sum of two terms, one containing $q^{2}_{_{1}} + q^{2}_{_{2}}$ and the other containing $q_{_{1}} q_{_{2}}$. This time the former term is \emph{not} neglected, reaching the conclusion that the perturbative approach for $a$ is unreliable. Crucially, however, the divergence is entirely due to the $q^{2}_{_{1}} + q^{2}_{_{2}}$ term: if we neglected it, we would remain with
\begin{equation}
a^{_{(2)\!}}(\La) = 1 - g^{2} \, \frac{q_{_{1}} q_{_{2}}}{2 V} \sum_{\veck \in \mcalR_{_{\!\La}}} \! \frac{\cos \Big[ \veck \cdot \big( \vecy_{_{\!1\!}} - \vecy_{_{\!2\!}} \big) \Big]}{\o_{k}^{3}} \quad ,
\end{equation}
which has a \emph{f\mbox{}inite} limit when we remove the cut-of\mbox{}f (again, if the summation is understood suitably, since the convergence is not absolute). The conclusion that evaluating $a$ with the perturbative approach we would get a series of the form $1 - g^{2} \cdot \infty + g^{4} \cdot \infty - \ldots$ is therefore entirely based on the decision to keep the $q^{2}_{_{1}} + q^{2}_{_{2}}$ term. Note that a f\mbox{}inite limit of $a^{_{(2)\!}}$ when removing the cut-of\mbox{}f would mean that the free and interacting spaces of state vectors are \emph{not} orthogonal.

It seems to us that the disparity is striking: the same type of term is neglected in one case and kept in the other, and this choice determines the f\mbox{}initeness or the divergence after removing the cut-of\mbox{}f. Neglecting the $q^{2}_{_{1}} + q^{2}_{_{2}}$ term in the f\mbox{}irst case implements the act of treating $B$ and $C$ dif\mbox{}ferently. If we treated $B$ and $C$ on the same ground, we would have an inf\mbox{}inite ground state energy (i.e.\ an ill-def\mbox{}ined model) and orthogonality of the free and interacting spaces of state vectors of the Hamiltonian (\ref{Fluo 1})--(\ref{Fluo 2}) (which could not be said, therefore, to describe the model). If we insisted in neglecting $C$, we should at least recognize that the diverging term in the expression for $a$ is related to it, and so for consistency we should neglect it as well. We would then have a f\mbox{}inite ground state energy and the orthogonality would disappear. This is exactly the moral that the study of the model with distributed sources teaches.

\subsubsection{The moral on the f\mbox{}irst part}

All considered, we are of the opinion that the Hamiltonian (\ref{Fluo 1})--(\ref{Fluo 2}) cannot be regarded as the quantum description of a scalar f\mbox{}ield interacting linearly with f\mbox{}ixed sources. For, if these source were point-like, then the condition (\ref{Fluo 2}) would hold but in this case we should retain the (divergent) $C$ in the Hamiltonian, and so (\ref{Fluo 1}) would not be correct. If on the other hand the sources were distributed, we could safely neglect $C$ and work with the Hamiltonian (\ref{Fluo 1}) but we should consistently abandon (\ref{Fluo 2}) in favor of a converging series.

Therefore, we propose that Van Hove's model can be considered at best as a pictorial inspiration to motivate the interest in the Hamiltonian (\ref{Fluo 1})--(\ref{Fluo 2}), but that there is no strict connection between the two. Moreover, in light of the strong correlation between the divergence of $C$ and the phenomenon of orthogonality of the spaces of state vectors, we believe that the latter phenomenon cannot be ascribed to a quantum scalar f\mbox{}ield interacting linearly with sources lying at f\mbox{}ixed positions (but just to the abstract Hamiltonian (\ref{Fluo 1})--(\ref{Fluo 2}) itself).

\subsection{The quantum Hamiltonian and its relevance}

The phenomenology associated to the Hamiltonian (\ref{Fluo 1})--(\ref{Fluo 2}) informs all the discussion in \cite{Van Hove 1952} regarding the formal structure of interacting QFT and the use of the perturbative approach in that context. Since this is the core and the most important part of Van Hove's article, the criticism we exposed in the previous section calls into question the relevance of Van Hove's article as a whole (save for its historical interest). Let us brief\mbox{}ly assess how Van Hove's discussion is af\mbox{}fected by our results.

\subsubsection{Van Hove's and our results}

Regarding the formal structure of QFT, the message that Van Hove's article transmits to the contemporary reader is that the use of UIR of the CCR is an important ingredient to describe an interacting QFT. Regarding the perturbative approach, instead, its main argument is that a change in the value of the coupling constant, no matter how small, changes the free space of state vectors into an altogether dif\mbox{}ferent space, and this should be taken into account when performing a perturbative analysis (with particular relevance for the interaction picture). Both of these arguments rest on the property of orthogonality of the spaces of state vectors relative to dif\mbox{}ferent values of the parameters of the Hamiltonian (\ref{Fluo 1})--(\ref{Fluo 2}).

It is fair to say that, if the orthogonality phenomenon were displayed just by an abstract Hamiltonian chosen ad hoc, one may doubt whether this really informs something about a realistic QFT. Conversely, if the Hamiltonian (\ref{Fluo 1})--(\ref{Fluo 2}) described the consistent quantization of a concrete (although crude) model of interacting f\mbox{}ield, the outlook of his results would be more prominent. From this point of view, the lack of strict connection between the Hamiltonian (\ref{Fluo 1})--(\ref{Fluo 2}) and Van Hove's model seem to subtract a fair amount of appeal from his article. For, to support the conjecture that the use of UIR of the CCR is important to describe an interacting QFT, it would now be necessary to motivate directly why the Hamiltonian (\ref{Fluo 1})--(\ref{Fluo 2}) should be representative of the formal structure of a realistic QFT. This was not done in Van Hove's article, and it is far from obvious why it should be so.

It is however worthwhile to point out that the model itself is very dif\mbox{}ferent from, and not easily related to, a realistic interacting QFT. Apart from the crudeness of the approximation on the structure of the sources, it is important to recall that non-linearities are absent in Van Hove's model: the quantum f\mbox{}ield is not self-interacting, and couples linearly with the sources. So it is far from clear what such a model could really teach about a realistic interacting case. From this point of view, the lack of strict connection between the Hamiltonian (\ref{Fluo 1})--(\ref{Fluo 2}) and Van Hove's model seems less tragic.

\subsubsection{The moral on the second part}

Overall, we believe that Van Hove's article retains some strength. From our point of view, its value does not lie in the certainty with which it indicates the mathematical properties which a non-perturbative QFT should have, but in the fact that it points to concepts, nowadays frequently overlooked, which is still worth keeping in mind. There is no certainty that these concepts will turn out to be useful. But, they should be known by those who, aiming to go beyond the perturbative renormalization prescription, look for a f\mbox{}inite and non-perturbative formulation of a realistic interacting QFT which be less lofty than the algebraic one \cite{Haag 1996} (if indeed this is possible). Other theoretical research programs may benef\mbox{}it by such a knowledge, see \cite{Thiemann Winkler 2001} for an example of application in Quantum Gravity.

Indeed Van Hove's article exposes very clearly how UIR of the CCR in a separable Hilbert space can coexist as embedded in an ``ambient'' non-separable Hilbert space, the inequivalence being ref\mbox{}lected in the orthogonality between distinct separable subspaces. It also shows that the appearance of UIR of the CCR is not to be ascribed just to very contrived and obscure Hamiltonians, but it emerges even from the simple Hamiltonian (\ref{Fluo 1})--(\ref{Fluo 2}). Moreover, it exposes these facts in a very clear way since the latter is exactly solvable. These are remarkable merits, which are unscathed by our present analysis.

On another point, it suggests that the very space of state vectors may not be something given once for all, but it may vary with the value of the coupling (more in general, with the Hamiltonian). Being aware of this possibility is important. The identif\mbox{}ication of the interaction picture as something problematic in QFT, potentially contributing to the appearance of perturbative divergences, is also a valuable insight which chronologically precedes and complements the body of results known under the name of ``Haag's theorem''.

If, in the intervening years from its publication, the awareness of these concepts had pervaded the physicists community, we could even say that Van Hove's article would retain just a historical interest. Since this has not happened, we are of the opinion that the article is still worth to be read.

\section{Conclusions}
\label{sec: Conclusions}

L\'{e}on Van Hove in \cite{Van Hove 1952} studied a simple model where a scalar quantum f\mbox{}ield interacts linearly with point-like masses lying at f\mbox{}ixed positions. He concluded that the spaces of state vectors relative to dif\mbox{}ferent values of the parameters of the Hamiltonian (such as the coupling constant, the number and the position of the sources) are orthogonal one to the other. While acknowledging the crudeness of the model, he conjectured that the spaces of state vectors relative to dif\mbox{}ferent values of the coupling constant would remain orthogonal, or at least dif\mbox{}ferent, also in a more realistic QFT. This conjecture was put forward despite the presence of a divergence in the exact solutions, which was discarded without any justif\mbox{}ication and whose physical meaning was not discussed.

We investigated the validity of Van Hove's conjecture by considering a slightly less irrealistic model where the sources still lie at f\mbox{}ixed positions, but have f\mbox{}inite spatial width. Our results indicate that, whenever the sources are modeled in a less irrealistic way, the phenomenon of orthogonality of the spaces of state vectors disappears. Moreover, we uncovered the physical meaning of the above mentioned divergence and showed that it is intimately linked to the orthogonality phenomenon, to the ef\mbox{}fect that regularizing the divergence kills the orthogonality of the spaces of state vectors.

We suggest that Van Hove's article is still worth reading since it displays, through a very simple Hamiltonian which can be solved exactly, the presence of Unitarily Inequivalent Representations of the Canonical Commutation Relations, which are associated to dif\mbox{}ferent values of the parameters of the Hamiltonian. Importantly, it shows how all these representations can coexist in a single non-separable Hilbert space. Moreover, it points out a possible harmful role of the interaction picture in the case where the space of state vectors depends on the coupling. These are potentially useful insights, even for applications beyond pure Quantum Field Theory, that are usually overlooked nowadays.

However, the lack of connection of this Hamiltonian even with a very crude QFT model weakens the authority of the conjecture that the mathematical structures uncovered in the article should \emph{necessarily} be an important part of the non-perturbative description of a realistic interacting QFT.

It is interesting to point out that, shortly after the appearance of Van Hove's article, an article by A.\ Loinger appeared \cite{Loinger 1952}, in which the quantized dynamics of two strings connected by springs was studied. The phenomenon of orthogonality of the spaces of state vectors was found to be absent in the model, and Van Hove of\mbox{}fered a qualitative explanation for this absence which did not contradict the results of his article, as recounted in \cite{Loinger erratum 1953}. In retrospective, it is hard not to wonder whether his argument was really as sound as previously thought.

\bigskip

\section*{Acknowledgments}

The author acknowledges partial f\mbox{}inancial support from the Funda\c{c}\~{a}o de Amparo \`{a} Pesquisa do Estado do Rio de Janeiro (FAPERJ, Brazil) under the Programa de Apoio \`{a} Doc\^{e}ncia (PAPD) program.

\appendix

\section{Results about the inf\mbox{}inite sums}
\label{app: main result}

In this appendix we collect results about the inf\mbox{}inite sums in (\ref{B}) and (\ref{C}). As we see below, the character of convergence of all these sums can be traced back to that of the inf\mbox{}inite sum
\begin{equation} \label{app infinite sum}
\frac{1}{V} \, \sum_{\veck \in \mcalR} \, \frac{1}{\o_{k}^{2}} \, \babs{\ti{\r}(\veck)}^{2} \quad ,
\end{equation}
where $\ti{\r}(\veck)$ is the Fourier coef\mbox{}f\mbox{}icient of a function $\r$ periodic on the cubic lattice and compactly supported in its primitive cell $\mscrP\,$.

A word about the notation. In what follows, a triple $\a = (\a_{_{1}}, \a_{_{2}}, \a_{_{3}})$ of non-negative integers is called a tri-index, and the space of tri-indices is indicated with $I_{^{+}}^{3}\,$. We indicate
\begin{align}
\vecx^{\a} &= x_{_{1}}^{\a_{_{1}}} \, x_{_{2}}^{\a_{_{2}}} \, x_{_{3}}^{\a_{_{3}}} \quad , & \de^{\b} &= \de_{_{1}}^{\b_{_{1}}} \, \de_{_{2}}^{\b_{_{2}}} \, \de_{_{3}}^{\b_{_{3}}} \quad .
\end{align}
We moreover indicate with $C_{c}^{\infty}$ the set of smooth functions with compact support, and $\mscrS$ indicates the Schwartz space of rapidly decreasing functions which by def\mbox{}inition satisfy
\begin{equation} \label{app Schwartz}
\sup_{\vecx \in \mbbR^{3}} \babs{\vecx^{\a} \, \de^{\b} f} < \infty \quad  , \qquad \forall \, \a , \b \in I_{^{+}}^{3} \quad .
\end{equation}

\subsection{Convergence}
\label{app subsec convergence}

To understand the behavior of $\ti{\r}(\veck)$ when $k \to \infty\,$, it is useful to introduce an auxiliary function $r$ closely related to $\r\,$. While the latter is periodic on the lattice, and it can be thought as being def\mbox{}ined on the primitive cell $\mscrP$ and extended by periodicity, let us consider a non-periodic function $r : \mbbR^{3} \to \mbbR$ which coincides with $\r$ only on the primitive cell, that is
\begin{equation} \label{app: auxiliary function}
r(\vecx) =
\begin{cases}
\, \r(\vecx)& \text{if} \,\, \vecx \in \mscrP \quad , \\[2mm]
\, 0& \text{elsewhere} \quad .
\end{cases}
\end{equation}
Clearly, $r$ belongs to $C_{c}^{\infty}(\mbbR^{3})$ as a consequence of $\r$ belonging to $C_{c}^{\infty} (\mscrP)\,$. Def\mbox{}ining the Fourier transform of $r$ as
\begin{equation} \label{app r Fourier transform}
\ti{r}(\veck ) = \int_{\mbbR^{3}} \! r(\vecx \, ) \, e^{-i \veck \cdot \vecx} \, d^{3}x \quad ,
\end{equation}
the comparison of (\ref{rho Fourier transform}) with (\ref{app r Fourier transform}) reveals that, by the way we def\mbox{}ined $r$, we have
\begin{equation} \label{app tirho tir}
\ti{r}(\veck) = \ti{\r}(\veck) \quad , \qquad \forall \, \veck \in \mcalR \quad .
\end{equation}
In other words, the Fourier transform of $r$ evaluated at a vector $\veck$ of the reciprocal lattice coincides with the Fourier coef\mbox{}f\mbox{}icient of $\r$ evaluated at the same $\veck\,$. With this in mind, it is possible to work with $\ti{r}$ instead of with $\ti{\r}$ whenever convenient.

Recall now that the Fourier transform, as an operator, is a bijection of $\mscrS(\mbbR^{3})$ onto $\mscrS(\mbbR^{3})$ \cite{Reed Simon II}. Since $C_{c}^{\infty}(\mbbR^{3}) \subset \mscrS(\mbbR^{3})$ it follows that $r \in \mscrS(\mbbR^{3})\,$, and therefore $\ti{r} \in \mscrS(\mbbR^{3})$ as well. By the property (\ref{app Schwartz}), the fact that $\ti{r}$ is rapidly decreasing implies that for every $\a \in I_{^{+}}^{3}$ we have
\begin{equation} \label{app emma 1a}
\sup_{\veck \in \mbbR^{3}} \, \bbabs{\, \veck^{\a} \, \frac{1}{\o_{k}^{2}} \, \babs{\ti{r}(\veck)}^{2}} < \infty \quad ,
\end{equation}
and taking into account (\ref{app tirho tir}) we get
\begin{equation}
\sup_{\veck \in \mcalR} \, \bbabs{\, \veck^{\a} \, \frac{1}{\o_{k}^{2}} \, \babs{\ti{\r}(\veck)}^{2}} < \infty \quad ,
\end{equation}
again for every $\a \in I_{^{+}}^{3}$. In particular, $\babs{\ti{\r}(\veck)}^{2}/\o_{k}^{2}$ goes to zero more rapidly than any non-negative power of $1/k\,$. Since the number of points of the reciprocal lattice contained in a sphere of radius $k$ diverge as a power of $k$ when $k \to \infty$, it follows that the inf\mbox{}inite sum (\ref{app infinite sum}) converges absolutely.

This result readily applies to the convergence of the inf\mbox{}inite sums in (\ref{C}), since both $\r_{_{1}\!}$ and $\r_{_{2}\!}$ are compactly supported. Regarding $(\ref{B})$, note that
\begin{multline}
\bbabs{ \, \Re \Big[ \, \ti{\r}^{\phantom{\ast}}_{_{1}}\!(\veck) \, \ti{\r}^{\ast}_{_{2}}(\veck) \, \Big] \, \cos \Big[ \veck \cdot \big( \vecy_{_{\!1\!}} - \vecy_{_{\!2\!}} \big) \Big] + \Im \Big[ \, \ti{\r}^{\phantom{\ast}}_{_{1}}\!(\veck) \, \ti{\r}^{\ast}_{_{2}}(\veck) \, \Big] \, \sin \Big[ \veck \cdot \big( \vecy_{_{\!1\!}} - \vecy_{_{\!2\!}} \big) \Big]} \leq \\[2mm]
\leq \bbabs{ \, \Re \Big[ \, \ti{\r}^{\phantom{\ast}}_{_{1}}\!(\veck) \, \ti{\r}^{\ast}_{_{2}}(\veck) \, \Big]} + \bbabs{\Im \Big[ \, \ti{\r}^{\phantom{\ast}}_{_{1}}\!(\veck) \, \ti{\r}^{\ast}_{_{2}}(\veck) \, \Big]} \leq 2 \, \babs{\ti{\r}^{\phantom{\ast}}_{_{1}}\!(\veck)} \, \babs{\ti{\r}^{\phantom{\ast}}_{_{2}}\!(\veck)} \quad ,
\end{multline}
and recall that the product of two rapidly decreasing functions is rapidly decreasing. It follows that the inf\mbox{}inite sum in (\ref{B}) converges absolutely.

\subsection{Inf\mbox{}inite volume limit}
\label{app subsec inf vol limit}

We now consider the inf\mbox{}inite volume limit
\begin{equation} \label{app infinite volume limit}
\lim_{V \to \infty} \, \frac{1}{V} \, \sum_{\veck \in \mcalR} \, \frac{1}{\o_{k}^{2}} \,\, \babs{\ti{\r}(\veck)}^{2} \quad .
\end{equation}

\subsubsection{Existence}
\label{subsubsec: Existence}

Note f\mbox{}irst of all that
\begin{equation} \label{Bianca}
\frac{1}{V} \, \sum_{\veck \in \mcalR} \, \frac{1}{\o_{k}^{2}} \,\, \babs{\ti{\r}(\veck)}^{2}  = \frac{1}{(2 \pi)^{3}} \,  \sum_{\veck \in \mcalR} \bigg( \frac{2 \pi}{L} \bigg)^{\!\! 3} \frac{1}{\o_{k}^{2}} \,\, \babs{\ti{r}(\veck)}^{2} \quad ,
\end{equation}
and that the inf\mbox{}inite sum on the right hand side is a Riemann sum of the integral
\begin{equation} \label{Tuane}
\int_{\mbbR^{3}} \frac{1}{\o_{k}^{2}} \, \, \babs{\ti{r}(\veck)}^{2} \, d^{3}k \quad .
\end{equation}
Specif\mbox{}ically, the cells of the reciprocal lattice are the 3-intervals of the Riemann sum.

The convergence of the integral (\ref{Tuane}) is easy to establish using the results of the previous section. In fact, writing the integral as the limit for $K \to \infty$ of the related integral on the sphere $\mcalB(K)$ of radius $K$ centered at the origin, we have
\begin{equation*}
\lim_{K \to \infty} \int_{\mcalB(K)} \frac{1}{\o_{k}^{2}} \, \, \babs{\ti{r}(\veck)}^{2} \, d^{3}k \leq \lim_{K \to \infty} \int_{0}^{K} \frac{4 \pi k^{2}}{\o_{k}^{2}} \max_{\norm{\vecz} = k} \babs{\ti{r}(\vecz)}^{2} dk \quad ,
\end{equation*}
and since $\ti{r} \in \mscrS(\mbbR^{3})$ the limit comparison test with the function $k^{-2}$ permits to positively conclude about the convergence. The expression (\ref{app infinite volume limit}) is therefore (proportional to) the limit of Riemann sums of a convergent integral, when the volume of the 3-intervals of the Riemann sums tend to zero. It follows that the inf\mbox{}inite volume limit (\ref{app infinite volume limit}) exists, and
\begin{equation} \label{Bridgette}
\lim_{V \to \infty} \frac{1}{V} \, \sum_{\veck \in \mcalR} \, \frac{1}{\o_{k}^{2}} \,\, \babs{\ti{\r}(\veck)}^{2} = \frac{1}{(2 \pi)^{3}} \int_{\mbbR^{3}} \frac{1}{\o_{k}^{2}} \, \, \babs{\ti{r}(\veck)}^{2} \, d^{3}k \quad .
\end{equation}

\subsubsection{Potential-mediated expression}
\label{subsubsec: Potential-mediated}

The expression on the right hand side of (\ref{Bridgette}) can be recast in a perhaps more intuitive form. Note that, since $r(\vecx) \, r(\vecz) \in C^{\infty}_{c}(\mbbR^{6})\,$, using Fubini's theorem and the def\mbox{}inition (\ref{app r Fourier transform}) we have
\begin{equation}
\babs{\ti{r}(\veck)}^{2} = \ti{r}^{\ast}(\veck) \, \ti{r}(\veck) = \iint_{\mbbR^{6}} \! r(\vecx) \, r(\vecz) \, \cos \Big[ \veck \cdot \big( \vecx - \vecz \big) \Big] \, d^{3}x \, d^{3}z \label{C 2} \quad ,
\end{equation}
so
\begin{equation} \label{eita}
\int_{\mbbR^{3}} \frac{1}{\o_{k}^{2}} \, \, \babs{\ti{r}(\veck)}^{2} \, d^{3}k = \int_{\mbbR^{3}} \frac{1}{\o_{k}^{2}} \, \, \Bigg\{ \iint_{\mbbR^{6}} \! r(\vecx) \, r(\vecz) \, \cos \Big[ \veck \cdot \big( \vecx - \vecz \big) \Big] \, d^{3}x \, d^{3}z \Bigg\} \, d^{3}k \quad .
\end{equation}
It would be attractive to express (\ref{eita}) as the integral in $d^{3}x \, d^{3}z$ of the product of the charge distributions mediated by a potential. As is well-known this is indeed possible, and it results into the appearance of the Yukawa potential mediating the interaction
\begin{equation} \label{miss Fluo}
\int_{\mbbR^{3}} \frac{1}{\o_{k}^{2}} \, \, \babs{\ti{r}(\veck)}^{2} \, d^{3}k = 2 \pi^{2} \!\! \iint_{\mbbR^{6}} r(\vecx) \, r (\vecz) \, \frac{e^{- m \norm{\vecx - \vecz}}}{\norm{\vecx - \vecz}} \, d^{3} x \, d^{3}z \quad .
\end{equation}

These relations allow us to express the numbers $B$ and $C$ in the forms (\ref{B inf vol}) and (\ref{C inf vol}). Regarding $C$, consider the expression (\ref{C}) and for each function $\r_{i}$ introduce the associate auxiliary function $r_{i}$ def\mbox{}ined as in section 
\ref{app subsec convergence}. The relation (\ref{Bridgette}) then implies that the inf\mbox{}inite volume limit of $C$ takes the form
\begin{equation}
C \xrightarrow[V \to \infty]{} - \frac{1}{2} \, \sum_{i = 1}^{2} \,\, \frac{g_{i}^{2}}{(2 \pi)^{3}} \int_{\mbbR^{3}} \frac{1}{\o_{k}^{2}} \, \, \babs{\ti{r}_{i}(\veck)}^{2} \, d^{3}k \quad ,
\end{equation}
and using (\ref{miss Fluo}) we arrive at
\begin{equation} \label{app C inf vol}
C \xrightarrow[V \to \infty]{} - \frac{1}{2} \, \sum_{i = 1}^{2} \,\, g_{i}^{2} \iint_{\mbbR^{6}} r_{i}(\vecx) \, r_{i}(\vecz) \, \frac{1}{4 \pi} \, \frac{e^{- m \norm{\vecx - \vecz}}}{\norm{\vecx - \vecz}} \, d^{3} x \, d^{3}z \quad ,
\end{equation}
which is (\ref{C inf vol}).

Regarding $B$, note that (\ref{B}) can be equivalently written as 
\begin{equation}
B = - \frac{g_{_{1}} g_{_{2}}}{V} \, \sum_{\veck \in \mcalR} \, \frac{1}{\o_{k}^{2}} \, \Re \bigg[ \, \ti{\r}^{\phantom{\ast}}_{_{1}}\!(\veck) \, \ti{\r}^{\ast}_{_{2}}(\veck) \, e^{-i \veck \cdot \big( \vecy_{_{\!1\!}} - \vecy_{_{\!2\!}} \big)} \bigg] \quad ,
\end{equation}
and that the same arguments of section \ref{subsubsec: Existence} can be used to infer that its inf\mbox{}inite volume limit exists and takes the form
\begin{equation} \label{MH}
B \xrightarrow[V \to \infty]{} - \frac{g_{_{1}} g_{_{2}}}{(2 \pi)^{3}} \, \int_{\mbbR^{3}} \frac{1}{\o_{k}^{2}} \, \Re \bigg[ \, \ti{r}^{\phantom{\ast}}_{_{\!1}}\!\!(\veck) \, \ti{r}^{\ast}_{_{\!2}}(\veck) \, e^{-i \veck \cdot \big( \vecy_{_{\!1\!}} - \vecy_{_{\!2\!}} \big)} \bigg] \, d^{3}k \quad .
\end{equation}
Furthermore, it is straightforward to verify that
\begin{multline}
\Re \bigg[ \, \ti{r}^{\phantom{\ast}}_{_{\!1}}\!\!(\veck) \, \ti{r}^{\ast}_{_{\!2}}(\veck) \, e^{-i \veck \cdot \big( \vecy_{_{\!1\!}} - \vecy_{_{\!2\!}} \big)} \bigg] = \\
= \iint_{\mbbR^{6}} r_{_{\!1}} (\vecx - \vecy_{_{\!1\!}}) \, r_{_{\!2}} (\vecz - \vecy_{_{\!2\!}}) \, \cos \Big[ \veck \cdot \big( \vecx - \vecz \big) \Big] \, d^{3} x \, d^{3} z \quad ,
\end{multline}
so the integral in (\ref{MH}) can be written as
\begin{equation} \label{eita p}
\int_{\mbbR^{3}} \frac{1}{\o_{k}^{2}} \,\, \Bigg\{ \iint_{\mbbR^{6}} \! r_{_{\!1}} (\vecx - \vecy_{_{\!1\!}}) \, r_{_{\!2}} (\vecz - \vecy_{_{\!2\!}}) \, \cos \Big[ \veck \cdot \big( \vecx - \vecz \big) \Big] \, d^{3}x \, d^{3}z \Bigg\} \, d^{3}k \quad .
\end{equation}
Analogously to the passage from (\ref{eita}) to (\ref{miss Fluo}), we have
\begin{multline}
\int_{\mbbR^{3}} \frac{1}{\o_{k}^{2}} \,\, \Bigg\{ \iint_{\mbbR^{6}} \! r_{_{\!1}} (\vecx - \vecy_{_{\!1\!}}) \, r_{_{\!2}} (\vecz - \vecy_{_{\!2\!}}) \, \cos \Big[ \veck \cdot \big( \vecx - \vecz \big) \Big] \, d^{3}x \, d^{3}z \Bigg\} \, d^{3}k = \\
= 2 \pi^{2} \!\! \iint_{\mbbR^{6}} r_{_{\!1}} (\vecx - \vecy_{_{\!1\!}}) \, r_{_{\!2}} (\vecz - \vecy_{_{\!2\!}}) \, \frac{e^{- m \norm{\vecx - \vecz}}}{\norm{\vecx - \vecz}} \, d^{3} x \, d^{3}z \quad ,
\end{multline}
so we get
\begin{equation} \label{app B inf vol}
B \xrightarrow[V \to \infty]{} - g_{_{1}} g_{_{2}} \iint_{\mbbR^{6}} r_{_{\!1}} (\vecx - \vecy_{_{\!1\!}}) \, r_{_{\!2}} (\vecz - \vecy_{_{\!2\!}}) \, \frac{1}{4 \pi} \, \frac{e^{- m \norm{\vecx - \vecz}}}{\norm{\vecx - \vecz}} \, d^{3} x \, d^{3}z \quad ,
\end{equation}
which is (\ref{B inf vol}).

\end{document}